# Temperature dependence of magnetic anisotropy in bulk and nanoparticles of $Pr_{0.5}Sr_{0.5}MnO_3$

## S.S.Rao and S.V.Bhat[*]

Department of Physics, Indian Institute of Science, Bangalore-560012, India

Nanoparticles (size 20, 40 and 60 nm) of $Pr_{0.5}Sr_{0.5}MnO_3$ are prepared by sol-gel technique and their magnetic properties are studied using ferromagnetic resonance and magnetization measurements. A comparison with the properties of the bulk material shows that the ferromagnetic transition at 265 K remains unaffected but the anti-ferromagnetic transition at $T_N$ = 150 K disappears in the nanoparticles. Further, the temperature dependence of magnetic anisotropy shows a complex behavior, being higher in the nanoparticles at high temperatures and lower at lower temperatures in comparison with the bulk.

[*]svbhat@physics.iisc.ernet.in ; for correspondence

# Introduction:

Nanoparticles and nanowires of doped rare-earth manganites [1-4] have recently been shown to exhibit strikingly different phases and properties compared to those of their bulk counterparts. For example, we have recently shown that there is either a weakening or a complete suppression of the charge ordered (CO) phase and a switch over from anti-ferromagnetic (AF) phase to ferromagnetic (FM) phase in nanowires of $Pr_{0.5}Ca_{0.5}MnO_3$ and $Pr_{0.57}Ca_{0.41}Ba_{0.02}MnO_3$ and nanoparticles of $Nd_{0.5}Ca_{0.5}MnO_3$ [5-7]. An issue still unresolved is the nature of magnetic anisotropy (MA) in the materials at nanoscale. For example, Shames et al observed that the MA is smaller in nanoparticles of $La_{0.9}Ca_{0.1}MnO_3$ compared to its value in the bulk [8]. This result is in conflict with the expectations and observations in literature, where it is reported that the MA in nanoparticles is larger than that in the bulk [9, 10]. Another aspect of MA that is being actively investigated presently is the temperature dependence of MA in nanoparticles. There is recent evidence [9] that MA in nanoparticles is a temperature dependent quantity even though it was earlier treated to be independent of temperature. MA happens to be an important parameter with respect to the application of magnetic materials for various magnetic data storage devices. Therefore, we have undertaken a detailed study of MA in doped rare earth manganites using ferromagnetic resonance (FMR) and magnetization measurements. In this report, we present our results on the bulk and nanoparticles of $Pr_{0.5}Sr_{0.5}MnO_3$ (PSMO). Our FMR results show that close to the FM transition temperature, MA in the bulk is considerably smaller than that of the nanoparticles. However, as the samples are cooled, it increases faster and becomes more than that in the nanoparticles below a certain temperature. Such a crossover in MA has not been reported so far to the best of our knowledge.

PSMO in the bulk is known to undergo a ferromagnetic transition at $T_C = 265$ K and when cooled further, transforms to an antiferromagnetic phase at $T_N = 150$ K [11, 12]. The composition dependence of MA in $Pr_{0.5+x}Sr_{0.5-x}MnO_3$ single crystals has recently been reported using torque magnetometry, magnetization and FMR measurements at room temperature [13]. It is found that even in the paramagnetic and AFM phases, there is evidence for the presence of MA indicating the occurrence of phase separation. However, a quantitative study of the temperature dependence of MA and the comparison of FMR results with those of magnetization are yet to be carried out. Here we report such a study for the bulk and nanoparticles (sizes 20, 40 and 60 nm) of PSMO.

## Experimental results and discussion:

The PSMO nanoparticles of size 20 nm were prepared by polymer assisted sol-gel method [7]. The as prepared nanoparticles were heated further to obtain particles of larger (40 and 60 nm) sizes. PSMO bulk is prepared by crushing single crystals of PSMO grown using float zone technique. X-ray diffraction and transmission electron microscopy are used to examine the phase purity, particle size and crystalline nature of the nanoparticles. The Rietveld refined XRD pattern of PSMO 20 is shown in figure 1; the inset shows its TEM micrograph. From the XRD pattern it is found that PSMO 20 crystallizes in the orthorhombic phase in the space group Pbnm with the unit cell parameters a = 5. 44325 $A^o$, b = 5.44758 $A^o$, c = 7.72473 $A^o$ and volume V = 229.058 $A^{o3}$ ; The corresponding bulk values are a = 5.443 $A^o$, b = 5.423 $A^o$ and c = 7.644 $A^o$, V = 225.63 $A^{o3}$ [14].

Magnetization measurements were carried out using a commercial vibrating sample magnetometer (VSM) attached to a physical property measurement system (PPMS) working in the temperature range from 2 K to 300 K. Magnetic field could be swept between –9 T and + 9 T. Figure 2 (a) presents the results of magnetization measurements as a function of temperature on bulk and nano PSMO 20 in the temperature range 4 K – 300 K measured in the presence of a magnetic field of 100 G. Figure 2 (b) shows the isothermal magnetization vs magnetic field plots at 10 K. It is clear from figure 2(a) that the bulk PSMO undergoes a paramagnetic (PM) -FM transition at ~ 265 K and a broad AFM transition at ~150 K in conformity with earlier reports. In contrast, in the case of PSMO 20 it is seen that while the transition to the FM phase remains practically unchanged (approximately the same $T_c$, slightly broader width), the transition to the AFM phase has completely disappeared. The latter behavior is similar to our earlier reports on nanowires of PCMO [5] and nanoparticles of NCMO [7]. The magnetization of bulk PSMO is found to be linearly dependent on H and is found to be completely reversible as expected for an anti-ferromagnet. The nanoparticles, instead, show hysteresis and saturation at relatively low magnetic field values. The inset shows the hysteresis loop in an expanded view. The saturation magnetization of the nanoparticles is found to be considerably smaller (1.53 $\mu_B$/f.u, obtained at 10 K, 5T) than that of the bulk value (3.5 $\mu_B$/f.u) similar to the observations in other systems.

In figures 3 (a) and (b), we present M vs H plots at 186 K in PSMO bulk and PSMO20 respectively. The temperature was chosen such that proper comparison can be made between the results of the two samples (both are ferromagnetic at this temperature). Insets to the figures 3 (a) and 3 (b) show the hysteresis behavior on an expanded scale. Coercive fields were determined from these plots. Similar experiments were carried out at 200 K and also on nanoparticles of sizes 40 and 60 nm. The values of MA were estimated from the hysteresis loops like the ones depicted in the insets of figure 3. The shapes of the hysteresis loops indicated that the magnetic anisotropy is of uniaxial type and the axes of different grains are randomly oriented. For such a system, the uniaxial anisotropy field $H_u$ can be calculated from $H_u = 0.479 \, H_c$ [15]. The anisotropy fields for PSMO bulk and PSMO 20 were obtained by following this procedure.

FMR measurements were carried out using a Bruker EMX X-band ESR spectrometer by sweeping the magnetic fields from 0 – 14900 G on loosely packed freestanding particles. DPPH was used as the field marker. In figure 4, the FMR signals are presented for bulk and nano PSMO20 at 186 K and 200 K. For the bulk sample, the signals are characteristic of uniaxial anisotropy. In the signals from the nano sample, due to the broadening caused by the disorder, the two features corresponding to $H_{r\parallel}$ and $H_{r\perp}$, where $H_{r\parallel}$ and $H_{r\perp}$ are the resonance fields for H parallel and perpendicular to the anisotropy axis, are not resolved. Nevertheless, the values of $H_{r\parallel}$ and $H_{r\perp}$ can be measured, and are used to estimate the uniaxial anisotropy using the formula [16] $H_{u(FMR)}$ = 2/3 ($H_{r\perp}$ - $H_{r\parallel}$). The values of uniaxial anisotropy thus obtained as a function of temperature for the bulk as well as the nanosized samples (20, 40 and 60 nm) are presented in figure 5. It is seen that the nano samples show much weaker temperature dependence of $H_{u(FMR)}$ compared to that of the bulk sample. The values of Hu for 40 and 60 nm particles are nearly identical and are larger than those for the 20 nm particle throughout the temperature range. The bulk values are smaller than those for nanoparticles for T > 195 K. Around this temperature, there is a cross over in the values, below which temperature the bulk values are higher than those of the nano particles. The bulk MA is seen to go through a peak around 185 K below which it decreases, most probably due to the impending antiferromagnetic transition.

As mentioned earlier [8], in nanoparticles of LCMO, MA was reported to be smaller than that of the bulk. However, in a number of reports [9, 10], MA for nanoparticles is shown to be larger than in the bulk. This is also expected on theoretical grounds since, due to the surface disorder the symmetry of small clusters is lower than that predicted from the crystal structure of the bulk. In fact, a random anisotropy analysis leads to the scaling of the contribution of the surface anisotropy as $1/R^2$, where R = average cluster radius. Though this is strictly true for cubic anisotropy, qualitatively similar conclusions can be drawn for uniaxial anisotropy as well.

The results reported in this work point towards a more complex scenario where the variation of relative magnitudes of anisotropy in nano and bulk samples as a function of temperature is seen to be quite different. As is seen from figure 5, the bulk $H_u$ is a strong function of temperature. For the nanosamples $H_u$ is found to increase with decreasing temperature though at a considerably lower rate than in the bulk. This property is expected to be useful for practical applications in magnetic data storage devices where magnetic anisotropy needs to be independent of temperature.

The anisotropy values estimated from the hysteresis plots at two temperatures (186 K and 200 K) for the bulk and the PSMO20 samples are also shown in the figure 5. These are seen to be significantly lower than the corresponding values obtained from FMR. Earlier reports have shown that the anisotropies determined from static measurements e.g., from coercive fields could be quite different from those at microwave frequencies [17,18]. This difference could arise either from surface relaxation effects or magneto elastic effects or both. All the same, for a sample of an assembly of grains with the randomly oriented axes of anisotropy, FMR is known to be a more reliable technique for the estimation of magnetic anisotropy.

**Conclusions:** In summary, we find that the antiferromagnetic phase observed in the bulk form of $Pr_{0.5}Sr_{0.5}MnO_3$ completely disappears in the nano-sized samples. The high temperature anisotropy of nanoparticles is much larger than that of the bulk sample. The MA of the nanosamples is a much less sensitive function of temperature than that of the bulk.

**Acknowledgements:** The authors gratefully acknowledge the help of Prof. C. N. R. Rao, Jawaharlal Nehru Centre for Advanced Scientific Research, Bangalore for providing the facilities for magnetization measurements. The Department of Science and Technology and the Council of Scientific and Industrial Research are thanked for financial support.

**Figure Captions:**

1. Figure 1 shows the observed (dots) and Rietveld fitted (continuous lines) XRD patterns of PSMO 20. The inset of this figure depicts the TEM micrograph of PSMO 20, non spherical and slightly elongated particles of diameter 20 nm are clearly seen.

2. Figure 2a) shows the variation of magnetization of PSMO 20 and PSMO BULK with the temperature (M-T) measured at 100 G while warming after the samples were cooled in zero field to the lowest temperature from room temperature (ZFC). Figure 2b) describes the isothermal magnetization variation with the magnetic field (M-H) ranges between –5T to + 5T at 10 K for both PSMO 20 and PSMO BULK. The inset of this figure shows the enlarged version of M-H of PSMO 20 measured at 10 K. The coercive field value is indicated as 930 G by an arrow.

3. Figures 3a) and 3b) show the M-H loops of PSMO BULK and PSMO20 at 186 K respectively. The expanded views of both figures are shown as insets in the respective figures, the coercive fields of PSMO BULK and PSMO20 are 81G and 181G are indicated in the inset figures by an arrow.

4. The FMR signals of PSMO BULK at 186 K and 200 K are shown in the figures 4a) and 4b) respectively. The sharp line which is seen at around 3300 G in these two figures is due to DPPH. Figures 4c) and 4d) show the FMR signals of PSMO20 at 190 K and 200 K respectively.

5. This figure shows the temperature variation of magnetic anisotropy ($H_u$) obtained from the FMR measurements performed on PSMO BULK (■), PSMO20 (●), PSMO40 (▲) and PSMO 60 (▼). This figure also contains the magnetic anisotropy values obtained from M-H measurements on PSMO bulk (◆) and PSMO 20 (◀) at 186 K and 200 K.

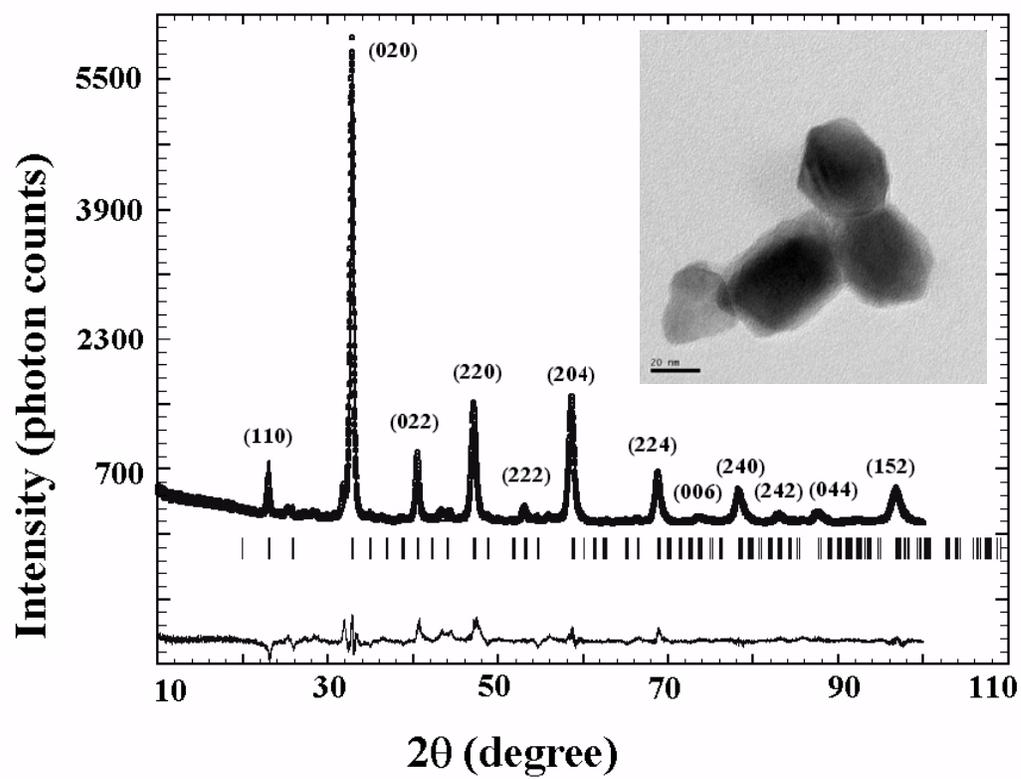

Figure 1  Rao et al.

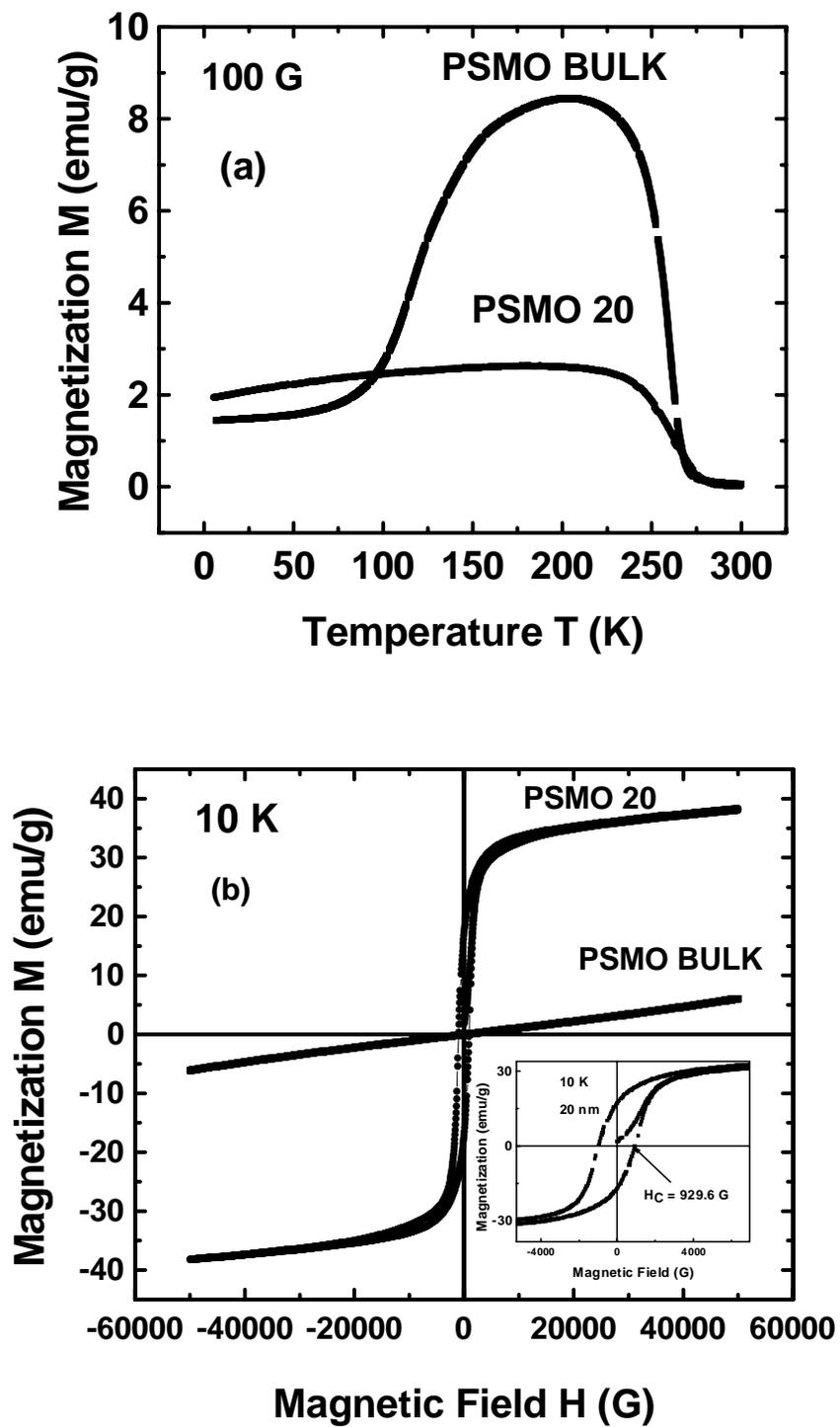

**Figure 2  Rao et al.**

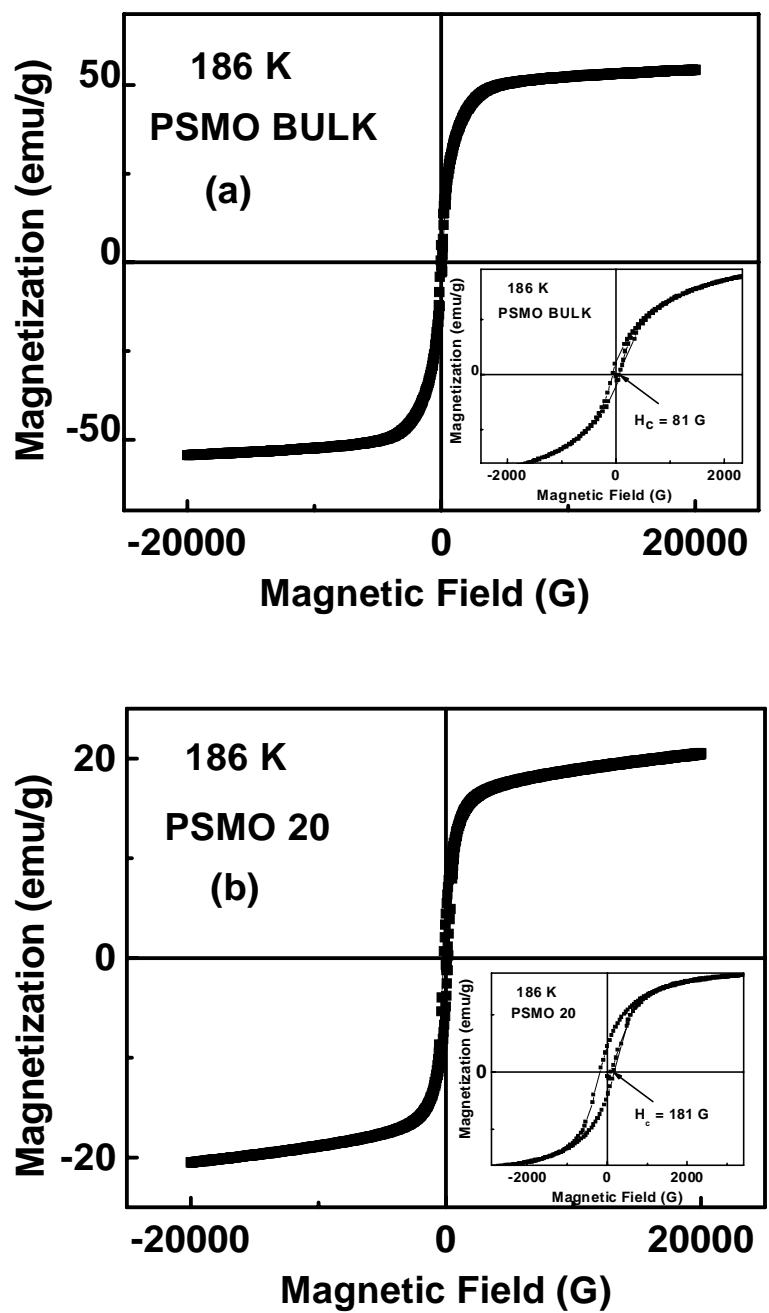

**Figure 3  Rao et al.**

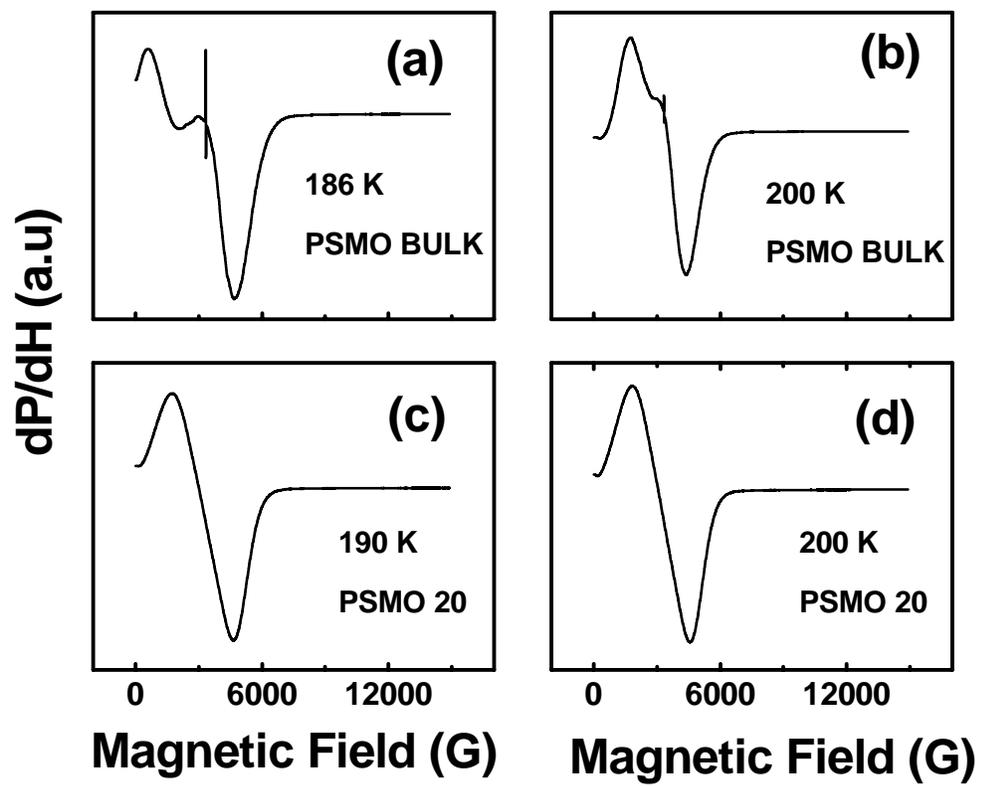

**Figure 4  Rao et al.**

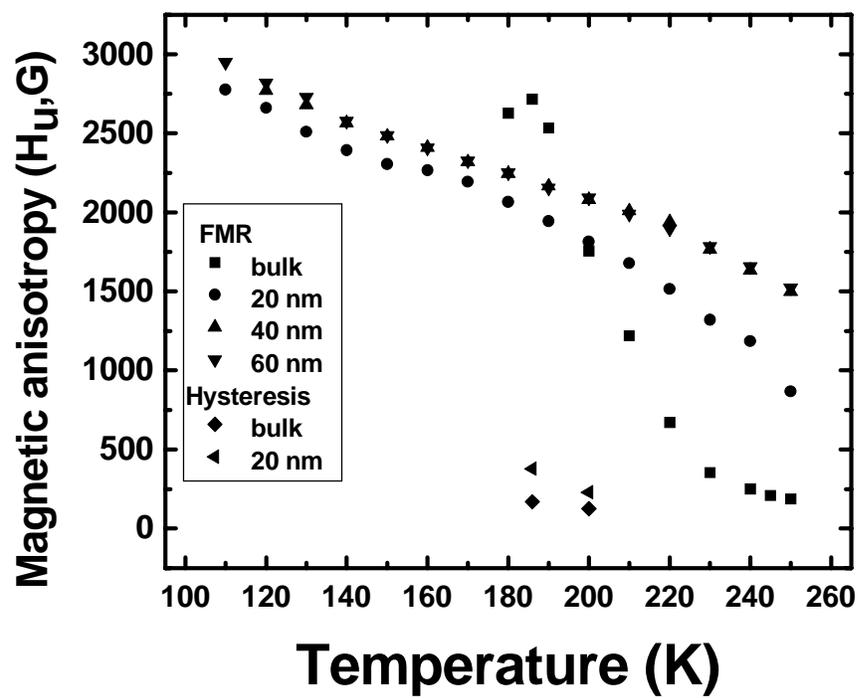

**Figure 5  Rao et al.**